\documentstyle[preprint,aps,tighten]{revtex}

\begin{document}
\draft
\title{Screened alpha decay in dense astrophysical plasmas and superstrong magnetic
fields}
\author{Theodore E. Liolios \thanks{%
www.liolios.info}}
\address{Hellenic Naval Academy of Hydra\\
School of Deck Officers, Department of Science\\
Hydra Island 18040, Greece}
\maketitle

\begin{abstract}
This paper shows that ultrastrong magnetic fields (such as those of
magnetars) and dense astrophysical plasmas can reduce the half life of alpha
decaying nuclei by many orders of magnitude. In such environments the
conventional Geiger-Nuttall law is modifed so that all relevant half lives
are shifted to dramatically lower values.
\end{abstract}


\section{Introduction}

Alpha radioactivity has been known for a long time in heavy nuclei. This
process, which is described in various textbooks of physics (see for example
Ref. \cite{gas}), can be described by the nuclear reaction

\begin{equation}
_{Z}^{A}X_{N}\rightarrow _{Z-2}^{A-4}X_{N-2}^{^{\prime }}+_{2}^{4}He_{2}+Q
\end{equation}
where we have used the usual textbook notation for the parent, the daughter
and the alpha nuclei. The value of the $Q$ energy, which is released during
the decay, can be derived by the application of the conservation-of-energy
principle which demands that

\begin{equation}
m_{X}c^{2}=m_{X^{^{\prime }}}c^{2}+m_{\alpha }c^{2}+Q_{n}
\end{equation}
where $m_{i}$ stands for the mass of nuclei $\left( i\right) \,$and $c\,\,$%
is the speed of light. This approach, adopted in most textbooks, yields the
energy released due to the rearrangement of nucleons which takes place
during the decay. Actually, we have used the subscript $\left( n\right) \,$%
in order to indicate that this energy is purely nuclear. However, if one
wants to be precise in his application of the conservation-of-energy
principle then the atomic nature of the reactants should also be taken into
account. Thus, the actual $Q$ energy released in the emission process is
given by

\begin{equation}
Q=Q_{n}+B_{e}\left( Z\right) -B_{e}\left( Z-2\right) -B_{e}\left( 2\right)
\end{equation}
Where $B_{e}\left( Z\right) ,B_{e}\left( Z-2\right) ,\,B_{e}\left( 2\right)
\,$ are the total electron binding energies of the parent, the daughter and
the alpha atoms, respectively. Although in typical terrestrial conditions,
such atomic corrections are of little importance they still have to be taken
into account if an accurate experimental value of the half life $T_{1/2}$ of
the decay is to be obtained. On the other hand, in certain astrophysical
environments screened alpha decay can present spectacular properties which
have never been investigated before. One of the main features of this paper
is to investigate those properties and their possible implications on some
long standing theories. The layout of the paper is as follows:

In Sec. II we study the effects of the electron cloud when alpha decay
occurs in a usual terrestrial environment. We derive new formulas which,
unlike others, can also take into account the degree of ionization of the
atomic cloud. The new formulas agree perfectly well with other less
sophisticated ones. In Sec. III the parent nucleus is considered to be under
the influence of a superstrong magnetic field such as the one encountered in
magnetars. In Sec. IV we study the alpha decay of nuclei in a dense
astrophysical plasma where the s and r process take place. In Sec. V, the
usual Geiger-Nuttall law is modified appropriately for magnetars and dense
astrophysical plasmas. Finally, Sec. VI presents briefly the conclusions of
the present paper.

\section{Screened alpha decay in a terrestrial environment.}

Let us assume that the parent nucleus is fully ionized (unscreened). During
alpha decay, outside the range of the nuclear forces, the alpha particle $%
\left( _{2}^{4}He_{2}\,\right) \,\,$experiences only the repulsive Coulomb
potential of the daughter nucleus $\left( _{Z-2}^{A-4}X_{N-2}^{^{\prime
}}\right) \,$so that the interaction energy will be

\begin{equation}
V_{c}\left( r\right) =\frac{2\left( Z-2\right) e^{2}}{r}
\end{equation}
The maximum height of the barrier will of course be

\begin{equation}
V_{0}=\frac{2\left( Z-2\right) e^{2}}{R}
\end{equation}
where $R$ is the minimum distance between the daughter nucleus and the alpha
particle roughly given by

\begin{equation}
R=1.3\,\left[ \left( A-4\right) ^{1/3}+4^{1/3}\right] \,fm
\end{equation}

The alpha decay half-life $T_{1/2}^{NSC}\,$\thinspace of an uscreened heavy
nucleus is inversely proportional to the penetration factor $P\left(
E_{\alpha }\right) $ given by the WKB method:

\begin{equation}
P\left( E_{\alpha }\right) =\exp \left[ -\frac{2\sqrt{2\mu }}{\hbar }%
\int_{R}^{r_{c}\left( E_{\alpha }\right) }\sqrt{V_{c}\left( r\right)
-E_{\alpha }}dr\right]  \label{penwkb}
\end{equation}
where the kinetic energy of the $\alpha $-particle is

\begin{equation}
E_{\alpha }=\frac{A-4}{A}Q_{n}  \label{eq}
\end{equation}
and the classical turning point is given by

\begin{equation}
V\left( r_{c}\right) =E_{\alpha }
\end{equation}
We will now define the two major limits of our study whose validity and
plausibility has been firmly established\cite{lioliosnpa} in the study of
multielectron screening effects in astrophysical reactions. Namely, if the
nucleus belongs in a neutral atom we should distinguish two modes of decay:

{\bf The adiabatic limit} $\left( AL\right) ,\,\,$which assumes that the
atomic clouds around the daughter and the alpha nuclei rapidly adjusts
themselves during the decay so that all the participants of the decay
(parent, daughter, alpha) are always in a neutral atomic form.

{\bf The sudden limit} $\left( SL\right) ,\,$where we assume that throughout
the decaying process the atomic cloud of the parent nucleus remains
undisturbed so that the daughter nucleus is screened by the same cloud as
the parent one and the alpha particle is emitted fully ionized. Note that in
that limit the neutral daughter atom will be assumed to have $Z-2\,\,$%
electrons\thinspace so that the TF theory can be used.

In the adiabatic limit the kinetic energy $E_{\alpha }$ will be:

\begin{equation}
E_{\alpha }^{AL}=E_{\alpha }+U_{e}  \label{alkinen}
\end{equation}
where the energy shift will be given by

\begin{equation}
U_{e}=\left( \frac{A-4}{A}\right) \left[ B_{e}\left( Z\right) -B_{e}\left(
Z-2\right) -B_{e}\left( 2\right) \right]  \label{uebind}
\end{equation}
The energy shift is usually much smaller than the kinetic energy $E_{\alpha
}\,\,$imparted on the alpha particle due to the rearrangement of nucleons
and can be calculated in the framework of the Thomas-Fermi theory\cite
{lioliosnpa}. According to previous studies\cite{assen} we can always define
a screening enhancement factor (SEF) so that 
\begin{equation}
f_{\alpha }\left( Z,A,Q_{n}\right) =\frac{P^{SC}\left( E_{\alpha
}+U_{e}\right) }{P^{NSC}\left( E_{\alpha }\right) }\geq 1  \label{fex}
\end{equation}
where $P^{SC}\left( E_{\alpha }+U_{e}\right) $ is the screened penetration
factor and $P^{NSC}\left( E_{\alpha }\right) \,\,$is the unscreened one.
Note that the kinetic energy $E_{\alpha }$ in Eq. $\left( \ref{fex}\right) $
refers to the unscreened nucleus.

Since $T_{1/2}\sim P^{-1}\left( E_{\alpha }\right) \,$\thinspace we can
write for the screened $T_{1/2}^{SC}\left( Z,A,Q_{n}\right) $ and the
unscreened $T_{1/2}^{NSC}\left( Z,A,Q_{n}\right) \,$\thinspace half lives :

\begin{equation}
T_{1/2}^{SC}\left( Z,A,Q_{n}\right) =\frac{T_{1/2}^{NSC}\left(
Z,A,Q_{n}\right) }{f_{\alpha }\left( Z,A,Q_{n}\right) }
\end{equation}
On the other hand, if the screening energy shift $U_{e}\,\,$is much smaller
than the kinetic energy $E_{\alpha }\,\,$of the alpha particle then

\begin{equation}
f_{\alpha }\left( Z,A,Q_{n}\right) =\exp \left( \pi n\frac{U_{e}}{E_{\alpha }%
}\right)  \label{fal}
\end{equation}
The screened half life will therefore be given by

\begin{equation}
T_{1/2}^{SC}\left( Z,A,Q_{n}\right) =\exp \left( -\pi n\frac{U_{e}}{%
E_{\alpha }}\right) T_{1/2}^{NSC}\left( Z,A,Q_{n}\right)  \label{scfnos}
\end{equation}
where $n$ is the Sommerfeld parameter for the interaction between the
daughter and the helium nuclei.

Obviously, the screening effect reduces the half-life of the decaying
nucleus. This is of course as expected since the screening cloud reduces the
Coulomb barrier thus easing the way of the alpha particle out of the parent
nucleus. In Fig.1, we have drawn a simplified picture of the screened alpha
decay. According to that figure the Coulomb potential practically vanished
at distances further than three screening radii (see next sections and Ref. 
\cite{lioliosnpa}).

The SEFs for heavy nuclei has been calculated in a paper\cite{lioliosnpa}
dealing with astrophysical nuclear reactions experiments. Actually the
derived formulas are particularly relevant here since the atoms involved in
alpha decay are always multielectronic. We can easily adjust those formulas
appropriately in order to describe the relevant screening effect in alpha
decay:

{\it Sudden Limit}

\begin{equation}
f_{\alpha }^{SL}\left( E\right) \simeq \exp \left[ -\frac{3.856\left(
Z-2\right) ^{7/3}}{Q_{n}^{3/2}}\left( \frac{A}{A-4}\right) \left[ S\left(
q\right) +\frac{q}{x_{0}\left( q\right) }\right] \right]
\end{equation}
where the degree of ionization is defined by

\begin{equation}
q=1-\frac{electrons}{protons}
\end{equation}
and the quantities $S\left( q\right) ,\,x_{0}\left( q\right) \,$ are defined
in Ref.$\,\cite{lioliosnpa}$. For neutral atoms $q=0,\,x_{0}\left( 0\right)
<\infty \,$ and $S\left( 0\right) =-1.588\,$so that

\begin{equation}
f_{TF}^{SL}\left( E\right) \simeq \exp \left[ \frac{6.1233\left( Z-2\right)
^{7/3}}{Q_{n}^{3/2}}\left( \frac{A}{A-4}\right) \right]  \label{sefsln}
\end{equation}
Note that the quantity $Q_{n}$ is measured in $keVs\,$\thinspace throughout
this paper$.$

{\it Adiabatic Limit}

\begin{equation}
f_{TF}^{AL}\left( E\right) \simeq \exp \left[ -\frac{62\left( Z-2\right)
\left[ F\left( q_{12}F\right) Z^{7/3}-F_{1}\left( q_{1}\right) \left(
Z-2\right) ^{7/3}-F_{2}\left( q_{2}\right) 2^{7/3}\right] }{Q_{n}^{3/2}}%
\frac{A}{A-4}\right]  \label{sefal}
\end{equation}
where\thinspace $F\left( q\right) \,$is defined in Ref. \cite{tal}:

\begin{equation}
F\left( q\right) =\frac{12}{7}\left( \frac{2}{9\pi ^{2}}\right) ^{1/3}\frac{%
e^{2}}{a_{H}}\left[ S\left( q\right) +\frac{q^{2}}{x_{0}\left( q\right) }%
\right]  \label{fq}
\end{equation}
and $a_{H}$ is the Bohr radius.

If we assume, according to the AL, that the parent, the daughter and the
alpha nuclei are all in a neutral atomic state then $F\left( q_{12}\right)
=F\left( q_{1}\right) =F\left( q_{2}\right) =-20.98\,\,eV\,\,$and the
relevant SEF is written:

\begin{equation}
f_{\alpha }^{AL}\left( E\right) \simeq \exp \left[ 1.297\left( Z-2\right) 
\frac{\left[ Z^{7/3}-\left( Z-2\right) ^{7/3}-2^{7/3}\right] }{Q_{n}^{3/2}}%
\frac{A}{A-4}\right]  \label{sefaln}
\end{equation}
We have compared Eq. $\left( \ref{sefsln}\right) $ to Eq. $\left( \ref
{sefaln}\right) $ and have found that their results practically coincide for
all alpha decaying nuclei. This remarkable coincidence proves the validity
of the present method and allows us to use the simple SL formula for the
description of the screening effect in terrestrial alpha decay.

Usually in alpha decay studies experimentalists use a semi-empirical formula
for screening energy

\begin{equation}
U_{e}=65.3\left( Z-2\right) ^{7/5}-80\left( Z-2\right) ^{2/5}\,eV
\end{equation}
which when inserted in Eq. $\left( \ref{scfnos}\right) $ gives roughly the
same results as Eq. $\left( \ref{sefsln}\right) .\,$

Thus we have derived alternative formulas for the accurate description of
the screening effect in alpha decay. Those formulas, which are based on the
solid mathematical framework of the TF theory, are the only ones available
that can take into account the degree of ionization of the participant
nuclei.

\section{Magnetically catalyzed alpha decay in magnetars}

Nowadays, there is a growing body of evidence (see Ref.\cite{pavlov} for a
review) for a population of neutron stars with magnetic fields of order $%
10^{15}G\,$\thinspace ,\thinspace which is much larger than the typical$\,$%
magnetic field of a neutron star (i.e. $10^{12}G)$. These ``magnetars'' are
distinguished from radio pulsars and accreeting binary neutron stars not
only by the strength of their field but also by the fact that their decaying
magnetic field is their primary energy source. Moreover, recent observations%
\cite{pavlov} provide strong evidence for the validity of the old hypothesis
that two separate classes of astronomical X-ray sources -- the Soft Gamma
Repeaters (SGR) and the Anomalous X-ray Pulsars (AXP)-- are actually
different manifestations of this peculiar type of star. The giant magnetic
field of magnetars has a significant and observable effect on quantum
electrodynamic processes operating near the star. It can also support strong
and persistent electrical currents, which alter the spindown of the star and
contribute to the continuous glow of X-rays and optical light observed in
between outbursts. In this section we will investigate its effects on the
abundances of alpha-decaying heavy elements which may find themselves in the
neighborhood of a magnetar.

In large magnetic fields, such as those existing in the atmospheres of
neutron stars, atomic clouds are compressed both perpendicular and parallel
to the magnetic field direction\cite{tal}. The effects of giant magnetic
fields $\left( B\geq 10^{12}\,G\right) \,\,$on hydrogen and helium atoms
have been extensively studied by many authors. Various studies have appeared
focusing on such topics as the formation of molecules and chains\cite{lai}
(and references therein) and nuclear fusion\cite{heyl}. However, no author
has ever considered the effects of such a magnetic field on alpha decay
processes.

Let us consider the heavy neutral atom of an alpha decaying element which is
under the influence of such an ultrastrong magnetic field. We will disregard
all exchange, thermal and relativistic effects as a first approximation and
adopt the usual super-magnetic field notation\cite{lai} $B_{12}=\left(
B/10^{12}\right) \,G,B_{0}=2.351\times 10^{9}G,\,b=B/B_{0}.$ Moreover, the
parent and the daughter nuclei are considered spinless (e.g. U-238,Th-234),
just like the alpha particle, so that we can disregard any coupling with the
external magnetic field. Note that the effect of a superstrong magnetic
field on nuclear properties has also been disregarded in the study of
magnetically catalyzed fusion reactions \cite{lai,heyl}. However, in such
cases where the fusing nuclei are not always spinless, coupling effects may
play a non-negligible role.

In any case the present study will exclusively focus on the perturbation of
half lives due to atomic (tunneling) effects allowing for an extra
perturbation term due to purely nuclear effects. This assumption is based on
the Born-Oppenheimer (BO) approximation according to which there is a
complete decoupling between electronic and nuclear degrees of freedom. (The
BO\ approximation has been used frequently in screening studies \cite
{bencze,langtr,bluge}).

{\it Sudden Limit}

The magnetic TF screened Coulomb potential will be given by

\begin{equation}
\Phi _{sc}\left( r\right) =\frac{Ze}{r}\phi \left( \frac{r}{R_{B}}\right)
\label{tfpotb}
\end{equation}
where the scaling parameter is $R_{B}=55133Z^{1/5}b^{-2/5}fm\,$ and the
universal function $\phi \left( x\right) \,\,$is given by Kadomtsev's\cite
{kadom} differential equation\thinspace with the initial conditions of Ref.%
\cite{banerjee}

\begin{equation}
\frac{d^{2}\phi \left( x\right) }{dx^{2}}=\left( x\phi \right) ^{1/2},\,\phi
\left( 0\right) =1,\,\phi ^{^{\prime }}\left( 0\right) =-0.938965
\end{equation}
where we have set $x=r/R_{B}\,.$

The above model is valid for neutral atoms when the condition $Z^{4/3}\ll
b\ll 2Z^{3}\,($or according to another study\cite{mueller}\thinspace $%
Z^{4/3}\ll b\ll 4.25Z^{3}\,)$ is satisfied.

In the sudden limit approximation the alpha particle, on its way out of the
parent nucleus, will have to penetrate the screened Coulomb potential given
by Eq. $\left( \ref{tfpotb}\right) $ so that the tunnelling will involve an
interaction potential energy given by

\begin{equation}
V_{sc}\left( r,B\right) =\frac{2\left( Z-2\right) e^{2}}{r}\phi \left( \frac{%
r}{R_{B}}\right)
\end{equation}
where $R_{B}=55133\left( Z-2\right) ^{1/5}b^{-2/5}fm\,$ and the respective
SEF will of course be given by the screened versus the unscreened
penetration factor:

\begin{equation}
f_{\alpha }^{SL}\left( Z,A,B\right) =\exp \left[ -\frac{2\sqrt{2\mu }}{\hbar 
}\left( \int_{R}^{r_{c}\left( E_{\alpha },B\right) }\sqrt{V_{sc}\left(
r,B\right) -E_{\alpha }}dr-\int_{R}^{r_{c}\left( E_{\alpha }\right) }\sqrt{%
V_{c}\left( r\right) -E_{\alpha }}dr\right) \right]  \label{slsefb}
\end{equation}
where the classical turning point in the magnetized alpha-decay is given as
usual by

\begin{equation}
V_{sc}\left( r_{c},B\right) =E_{\alpha }
\end{equation}
We might follow the treatment of Sec.II where we derived approximate
analytic SL SEFs for conventional alpha-decay, assuming that there exists a
constant screening energy shift (much smaller than E$_{\alpha })$. This
method, which actually replaces Eq. $\left( \ref{slsefb}\right) $with Eq. $%
\left( \ref{fal}\right) ,\,$would indeed yield very elegant analytic SEFs
but we cannot afford to make any approximations yet. This is due to the fact
that we are studying a completely novel effect and thus we must be certain
about the accuracy of our results. Thus we will numerically evaluate the SL
SEFs given by Eq. $\left( \ref{slsefb}\right) $.

Moreover, in some cases, relativistic corrections to the TF atom may become
important. In order to investigate relativistic effects we will employ the
equation derived by Hill, Grout and March\cite{hill} and Shivamoggi and
Mulser\cite{shiv}

\begin{mathletters}
\begin{equation}
\frac{d^{2}\phi \left( x\right) }{dx^{2}}=\left( x\phi \right) ^{1/2}\left(
1+\Lambda \frac{\phi }{x}\right) ^{1/2},\,\phi \left( 0\right) =1,\,\phi
^{^{\prime }}\left( 0\right) =-0.938965  \label{tfrel}
\end{equation}
where the relativistic parameter $\Lambda $ stands for 
\end{mathletters}
\begin{equation}
\Lambda =\frac{Ze^{2}}{2m_{e}c^{2}R_{B}}\ll 1
\end{equation}
or else

\begin{equation}
Z^{4/5}b^{2/5}\ll 38286
\end{equation}

We have run extensive numerical integrations of Eq. $\left( \ref{slsefb}%
\right) $ applying the above relativistic model to various magnetic fields
and heavy nuclei. Provided that the conditions $\Lambda \ll 1\,$and $%
Z^{4/3}\ll b\ll 2Z^{3}\,\,$are valid, we have concluded that relativistic
corrections to $f_{\alpha }^{SL}\left( Z,A,B\right) $ are negligible.

{\it Adiabatic Limit}

In an ultrastrong magnetic field, due to the multielectron nature of an
alpha decaying atom, the sudden limit is expected to yield practically the
same results as the adiabatic limit. This has been shown in the previous
section for conventional screened alpha decay and common sense demands that
this is the case when supermagnetized atoms are considered. It is obvious
that subtracting two electrons from the large number of them which orbit the
parent nucleus will induce a very small perturbation to the charge
distribution around it. This of course means that the sudden limit is
expected to be very accurate just as was shown in the previous section. We
can use the total binding energy of a supermagnetized heavy atom $E\simeq
-13.6Z^{9/5}b^{2/5}\,eV$, in order to obtain the screening shift yielded by
the adiabatic limit which, according to Eq. $\left( \ref{uebind}\right)
,\,\, $reads

\begin{equation}
U_{TF}^{AL}=0.0136\,b^{2/5}\left[ Z^{9/5}-\left( Z-2\right)
^{9/5}-2^{9/5}\right] \,keV  \label{ualion}
\end{equation}
and after some algebra the relevant AL SEF is found to be given by the
formula

\begin{equation}
f_{\alpha }^{AL}\left( Z,A,Q,B\right) \simeq \exp \left[ \frac{0.85}{%
Q_{n}^{3/2}}\left( Z-2\right) \left( \frac{A}{A-4}\right) b^{2/5}\left[
Z^{9/5}-\left( Z-2\right) ^{9/5}-2^{9/5}\right] \right]  \label{alsef}
\end{equation}
In Fig.2 ,we have numerically integrated Eq. $\left( \ref{slsefb}\right) $%
\thinspace in order to plot the magnetic SL SEF for the alpha decay of $%
^{238}U\,\left( T_{1/2}=4.46\times 10^{9}\,y\right) ,\,$and $^{235}U\,$ $%
\left( T_{1/2}=0.7\times 10^{9}\,y\right) $with respect to the magnetic
field strength (measured in units of $2.351\times 10^{9}G).\,$We have also
included the AL SEFs given by Eq. $\left( \ref{alsef}\right) .$ The solid
vertical bar signifies the upper limit of our model for the nuclei in
question, while the lower limit is actually that field for which the SEF
becomes roughly unity. The results of both limits are very close to each
other just as predicted.

We have particularly chosen these two uranium isotopes as they are
thoroughly used as cosmochronological tools\cite{claytonura,goriely}. By
observing the reduction in the half lives of those alpha-decaying isotopes
in Fig.1 we can argue that ultrastrong magnetic fields act as giant
transformers of $^{238}U,^{235}U$ into their respective daughters $%
^{234}Th\, $and $^{231}Th$.

According to Fig.2, magnetars can reduce the half life of uranium by four
orders of magnitude. The effect is of a similar order of magnitude for other
heavy alpha decaying nuclei as well. Although the mathematics of our model
forbids its use at fields larger than $10^{15}G$ it is more than obvious
that half lives will be further reduced at ever stronger fields where our
model is invalid.

Another interesting fact about ultramagnetized alpha decay is that the
compression of the electron cloud is particularly large in the direction
perpendicular to the magnetic field, while it is very small in the parallel
direction. Thus, the emission of alpha particles will not be isotropic as is
usually the case in terrestrial process but it will occur in such a way that
it peaks in the perpendicular direction. The phenomenon of anisotropically
enhanced alpha-decay has never been investigated before. In Sec. IV we will
prove that it also appears in the s and r processes in stellar plasmas,
although in such sites the screened half lives can be up to 9 orders of
magnitude smaller than the unscreened ones.

Finally, we note that the screened half life (the SEF)\ is an increasing
(decreasing) function of the decay energy $Q_{n}\,.\,$This is due to the
fact that the classical turning point is a decreasing function of the energy 
$Q_{n}$ so that the smaller the $Q_{n}\,$the thicker the barrier that the
alpha particle will have to cross and thus the stronger the screening
effect. Our tests have shown that the TF screened Coulomb potential exhibits
a marked deviation from the unscreened one mainly at large distances from
the nucleus. Thus, large turning points allow the screening effect to play a
more important role in the tunnelling process.

\section{Screened alpha decay in dense astrophysical plasmas}

Although various authors have studied the effects of a very dense
astrophysical plasma on fusion reaction rates, no author has ever studied
such effects on the alpha decay process.

Actually, heavy nuclei which decay by alpha particle emission exist only in
the form of seeds in ordinary massive stars where the zero metallicity
scenario is usually valid for most stellar evolution calculations. For
example in Population I stars the uranium abundance is roughly\cite
{cameronbook} eleven (six) orders of magnitude smaller than that of hydrogen
(silicon). Nevertheless, there are stellar processes such as the s(low) and
r(apid) ones which generate a significant number of heavy nuclei which are
then ejected into space via a supernova explosion. Admittedly, the
production of such nuclei doesn't play any significant role in stellar
evolution which is governed by light element production-destruction
processes. However, the abundances of heavy elements give important
information about the formation of the universe and therefore all factors
which influence them deserve special attention. In this section we will
prove that alpha decay in dense stellar plasmas can play a much more
important role in the destruction of heavy elements than initially thought.

Let us consider a heavy alpha decaying nucleus $\,_{Z}^{A}M_{N}$ in a fully
ionized multicomponent plasma which is at thermodynamic equilibrium. We will
modify Mitler's model\cite{mitler} for screened thermonuclear reactions in
order to derive screening corrections in our alpha-decay study. Actually,
this modification is perfectly legitimate since all plasma screening models
are concerned with the perturbation of the penetration factor $P\left(
E\right) \,\,$which is the same for both fusion and decay.

{\it Sudden Limit}

In that limit we assume that the plasma which screens the nucleus $%
_{Z}^{A}M_{N}$ \thinspace remains undisturbed by the emission of the alpha
particle. According to Sec. II and Sec. III. we model this process by
considering the interaction between the daughter nucleus and the alpha
particle inside the plasma. If we modify Mitler's model the screened Coulomb
potential is given by

\begin{eqnarray}
V_{sc}^{M}\left( r\right) &=&\frac{2\left( Z-2\right) }{r}%
-C_{o}+C_{1}r^{2}\qquad \qquad r<r_{0} \\
V_{sc}^{M}\left( r\right) &=&\frac{2\left( Z-2\right) C}{r}\qquad \qquad
\qquad \qquad \qquad r>r_{0}
\end{eqnarray}
where

\begin{equation}
x=\frac{r_{0}}{R_{D}}=\left( \frac{3\left( Z-2\right) }{4\pi n_{e}R_{D}^{3}}%
+1\right) ^{1/3}-1
\end{equation}
$R_{D}$ is the usual Debye-Huckel radius (corrected of course for electron
degeneracy) , $n_{e}\,$is the average electron number density in the plasma
and the constants $C_{0},\,C_{1}$ are given by

\begin{equation}
C_{1}=\frac{2}{3}\pi en_{e}\qquad \qquad C_{0}=2\pi en_{e}R_{D}^{2}x\left(
x+2\right)
\end{equation}
In order to derive a simple analytic formula for the SL SEF let us first
assume that the screening energy due to the stellar plasma is much smaller
than the decay energy $Q_{n}$ , which is usually a few $MeVs$. To the extend
that this assumption is wrong then our calculation would yield a
conservative estimate of the associated SEF (i.e. the SEF will certainly be
larger). However, as we will prove, this assumption is perfectly legitimate
in most stellar plasmas away from solidification. In such a case the
screening energy will be the properly modified Mitler's shift:

\begin{equation}
U_{e}^{M}=\frac{2\left( Z-2\right) e^{2}}{R_{D}}g\left( x\right)
\end{equation}
where $g\left( x\right) $

\begin{equation}
g\left( x\right) =\left( \frac{1+x_{1}/2}{1+x_{1}+x_{1}^{2}/2}\right)
\end{equation}
then using Eq. $\left( \ref{fal}\right) $ we obtain

\[
f_{\alpha }^{M}=\exp \left( \pi n\frac{U_{e}^{M}}{Q_{n}}\right) 
\]
or else

\begin{equation}
f_{\alpha }^{M}\left( Z,A,\rho ,T\right) =\exp \left( \frac{2\left(
Z-2\right) e^{2}\pi ng\left( x\right) }{Q_{n}R_{D}}\right)  \label{fmad}
\end{equation}
{\it Adiabatic Limit}

In order to be more precise we have to take into account the screening
effects induced by the alpha particle as well as that the assumption of a
very small screening energy is not necessarily true for all cases. Both
those factors are taken into account by the adiabatic limit. If we further
assume that the stellar plasma where the alpha-decay takes place has not
reached the solidification point, which is the case in s and r process
environments, then the screening enhancement factor will be the respective
Mitler's\cite{mitler} SEF modified appropriately for an alpha-decay process:

\begin{equation}
f_{M}=\left( f_{S}\right) ^{g\left( \zeta _{1},\zeta _{2}\right) }
\label{sefm2}
\end{equation}
where $f_{S}$ is the usual Salpeter's\cite{salpeter} SEF and the parameter $%
g\,\,\,$is\cite{lioliosrp}

\begin{equation}
g\left( \zeta _{1},\zeta _{2}\right) =\frac{9}{10}\left( \frac{1}{\zeta
_{1}\zeta _{2}}\right) \left[ \left( \zeta _{1}+\zeta _{2}+1\right)
^{5/3}-\left( \zeta _{1}+1\right) ^{5/3}-\left( \zeta _{2}+1\right)
^{5/3}+1\right]
\end{equation}
where $\zeta _{1},\zeta _{2}$ are dimensionless parameters which for the
alpha decay process are given by

\begin{equation}
\zeta _{1}=\frac{3\left( Z-2\right) }{4\pi N_{e}R_{D}^{3}},\qquad \qquad
\zeta _{2}=\frac{3\times 2}{4\pi N_{e}R_{D}^{3}}  \label{zdms}
\end{equation}
Thus, the screened half-life of a particular alpha-decaying heavy nucleus
will now be a function of plasma composition\thinspace , density and
temperature:

\begin{equation}
T_{1/2}^{SC}\left( \rho ,T\right) =\left( f_{\alpha }^{M}\right)
^{-1}T_{1/2}^{NSC}
\end{equation}
We have compared the results given by Eq. $\left( \ref{sefm2}\right) $ and
Eq.$\left( \ref{fmad}\right) $ and have found that they are practically the
same for all relevant stellar environments.. Therefore the simple
formula\thinspace given by Eq. $\left( \ref{fmad}\right) $describes
accurately the reduction of the half-life of the screened alpha-decaying
nuclei. In any case, the SEF is well constrained by Eqs. $\left( \ref{fmad}%
\right) $ and $\left( \ref{sefm2}\right) .$

The whited out figures of Fig.3 represent the ratio of the unscreened
half-life $T_{1/2}^{NSC}\,$\thinspace versus the screened one $T_{1/2}^{SC}$%
\thinspace \thinspace (i.e. the SEF)\ for the isotope $^{238}U\,\,$in
various stellar environments. The screening reduction of half lives is not
very sensitive to temperature for completely degenerate environments.. This
is due to the fact that, as can be shown after some algebra, in such
ultradense environments the screening energy is approximately given for both
limits (see Fig.2) by the simple formula

\begin{equation}
U_{e}^{AL}=0.0176\,\left( \frac{\rho }{\mu _{e}}\right) ^{1/3}\left[
Z^{5/3}-\left( Z-2\right) ^{5/3}-2^{5/3}\right] \,keV  \label{ueal0}
\end{equation}
which is independent of temperature. Eq. $\left( \ref{ueal0}\right) \,$is
actually Salpeter's\cite{salpeter} formula for completely degenerate
electron gases modified appropriately for our study. The relevant SEF is of
course still given by Eq. $\left( \ref{fal}\right) $. Note that, according
to Fig.3, in supernova shocks, where the r process takes place, the
screening effect is particularly accentuated.

Most heavy nuclei, which undergo alpha decay, are produced\cite{hoyle}
during the s and r processes of stellar evolution either during a long epoch
of thousands of years or during short pulses and shocks of milliseconds. Let
us assume that such a nucleus of abundance $N\left( t\right) $ \thinspace is
produced in a dense stellar plasma . We know that this abundance will
actually follow the usual law of exponential decay that is 
\begin{equation}
N\left( t\right) =N\left( 0\right) \exp \left( -\frac{\ln 2}{T_{1/2}}t\right)
\end{equation}
According to the evolutionary stage of the star, there are various
mechanisms which generate or destroy the heavy nucleus in question with the
paramount ones being neutron capture (i.e. s and r processes), beta decay
and photodisintegration. It a very an important finding of the present paper
that alpha decay half lives in dense astrophysical plasmas can become so
small that alpha decay can play an equally significant role in the evolution
of heavy element abundances. Instead of presenting a detailed analysis of
this effect we can give a fair approximation to the actual extend of the new
effect by comparing the screened half lives to the timescale of the
destruction/production mechanisms: First we note that if we disregard all
other factors then alpha decay alone can reduce the stellar abundance of a
nucleus by three orders of magnitude within ten half lives. Since the half
life itself in a screened environment can be many orders of magnitude
smaller than the unscreened one it is obvious that a new important mechanism
of destruction has been discovered which so far has been considered
negligible for a lot of heavy elements. In fact if $\tau $ is the time scale
for a certain process which produces or destroys a heavy nucleus then the
abundances of all nuclei whose unscreened half life is of the order of $\,$

\begin{equation}
T_{1/2}^{NSC}\sim f_{\alpha }^{M}\left( Z,A,\rho ,T\right) \times \tau \,
\end{equation}
will be considerably affected by the alpha decay process . Considering that
the timescales of the s and r processes vary\cite{claytonbook} from seconds
to millions of years the importance of the present findings in now obvious.

\section{The Geiger-Nuttall law for magnetars and dense thermonuclear plasmas
}

The success of the quantum mechanical description of alpha decay has been
established by the Geiger-Nuttall\thinspace (GN) law\cite{geiger} which is
described in most textbooks dealing with alpha decay theory. According to
that law, in an unscreened environment, a good fit to the half life data $%
T_{1/2}$ of a large number of alpha emitters is obtained with the formula

\begin{equation}
\log _{10}T_{1/2}\left( Z,A,Q_{n}\right) =C_{1}\left( Z\right)
Q_{n}^{-1/2}+C_{2}
\end{equation}
where $C_{2}\,$is a constant and $C_{1}\left( Z\right) $ a slowly varying
parameter of the atomic number\thinspace $Z$.

These relationships have been proved more effective than most
microscopically based calculations in the prediction of alpha decay half
lives. Their application to the decays of all isotopic sequences of the
heaviest elements with neutron number $N>126$ has long been known\cite{galla}
to yield spectacularly straight line plots. The validity of this linear
correlation has been established\cite{tuli,browne} for lighter nuclei, as
well.

According to the new findings of the present paper the GN law in magnetar
atmospheres and dense thermonuclear plasmas should be modified. Therefore,
if the GN law in an unscreened environment is given as a plot of the half
life with respect to the atomic number and the decay energy then in the
previously studied screened environments all such plots should be modified
so that the readings on the mantissa should be shifted by $\log
_{10}f_{\alpha }^{-1}.\,$Thus, in our study of alpha decay in magnetars and
dense plasmas, we can use all conventional GN plots and data currently
available provided we apply the following rules:

\begin{equation}
\log _{10}T_{1/2}^{SC}\left( Z,A,B\right) =\log _{10}T_{1/2}^{NSC}\left(
Z,A\right) -\log _{10}f_{\alpha }^{TF}\left( Z,A,B\right)
\end{equation}
for magnetars and

\begin{equation}
\log _{10}T_{1/2}^{SC}\left( Z,A,\rho ,T\right) =\log
_{10}T_{1/2}^{NSC}\left( Z,A\right) -\log _{10}f_{\alpha }^{M}\left(
Z,A,\rho ,T\right)
\end{equation}
for dense stellar plasmas.

A final argument concerning heavy element production/destruction should be
expressed: Alpha decay is a nuclear process which bears a lot of physical
similarities to fission. Since fission is also important (e.g. the
californium hypothesis\cite{hoyle}) in the evolution of heavy element
abundances we argue that similar strong screening effects are bound to
appear when fissionable nuclei exist in the astrophysical environments
discussed in the present paper.

\section{Conclusions}

We have studied electron screening effects in alpha decay processes applying
a formalism which so far has been exclusively used in the study of
astrophysical fusion reactions. We have derived alternative analytic SEF
formulas for terrestrial alpha decay processes which can also take into
account the degree of ionization of the decaying atom.

More importantly, this paper also studies the effects of superstrong
magnetic fields (such as those of magnetars) on alpha decay proving that the
relevant half life can be reduced by several orders of magnitude. The whole
effect, which is expressed in the form of a very handy formula, namely Eq. $%
\left( \ref{alsef}\right) ,$ may possibly have notable implications on heavy
element abundances and the cosmochronological models that rely on them.

Finally, there has been shown, for the first time, that alpha decay half
lives in dense astrophysical plasmas can be reduced by many orders of
magnitude due to plasma screening. Those results may have significant
implications on the evolution of heavy element abundances during the s and r
processes. A very simple analytical formula has been produced (i.e. Eq.$%
\,\left( \ref{fmad}\right) $) which can take into account all those novel
effects.

FIGURE CAPTIONS

Figure 1. A simplified picture of screened alpha decay. The alpha particle
is emitted with a (relative) kinetic energy $E_{\alpha },\,$while the
screened $\left( r^{SC}\right) \,$and unscreened $\left( r^{NSC}\right) \,$%
classical turning points are also shown.\thinspace Note that the maximum
height of the Coulomb barrier in the screened case will be shifted downwards
by $\Delta V_{c}$ while the nuclear state of the parent nucleus is described
by a potential well of depth $-V_{0}.$ The relative distance is measured in
screening radii while the mantissa has been modified (exaggerated) in
certain points to help visualization of the effect.

Figure 2.The ratio of the unscreened half life $T_{1/2}^{NSC}\,$\thinspace
to the screened one $T_{1/2}^{SC}$\thinspace \thinspace (i.e. the SEF)\ for
two important alpha decaying isotopes with respect to the magnetic field
strength (measured in units of $2.351\times 10^{9}G):\,^{238}U\,$\thinspace
(upper/lower solid curves), $^{235}U$ (upper/lower dotted curves).\thinspace
The upper (lower) curves stand for the AL (SL) SEFs\thinspace for each
isotope$.\,$

Figure 3. The ratio of the unscreened half-life $T_{1/2}^{NSC}\,$\thinspace
to the screened one $T_{1/2}^{SC}$\thinspace \thinspace (i.e. the SEF)\ for
the isotope $^{238}U\,\,$in various stellar plasmas. The vertical column of
values $\left( -10,-5,\,etc.\right) \,\,$on the right-hand-side mantissa
stands for the well-known degeneracy parameter $a\,\,$which is related to
the electron chemical potential $\mu _{e}\,$via the formula $a=-\mu _{e}/kT$%
.\thinspace \thinspace In the plot five electron degeneracy regimes are
shown: ND:Non Degenerate, WD:Weakly Degenerate, ID:Intermediate Degeneracy,
SD:Strong Degeneracy, CD:Complete Degeneracy (defined in Ref. \cite
{lioliosnpa}). The numbers in the whited-out areas of the plot correspond to
the SEFs for $^{238}U\,\,\,$calculated according to the theory of Sec.IV.
Various stellar sites are shown while the thick horizontal line at $\rho
=7.3\times 10^{6}\,g/cm^{3}\,$defines the relativistic domain of the the
equation of state.

\medskip

{\it This work was presented during the conference ''Supernova, 10 years of
SN1993J'', April 2003, Valencia, Spain. The author is grateful to Prof.
Hillebrandt for useful comments and discussions.}

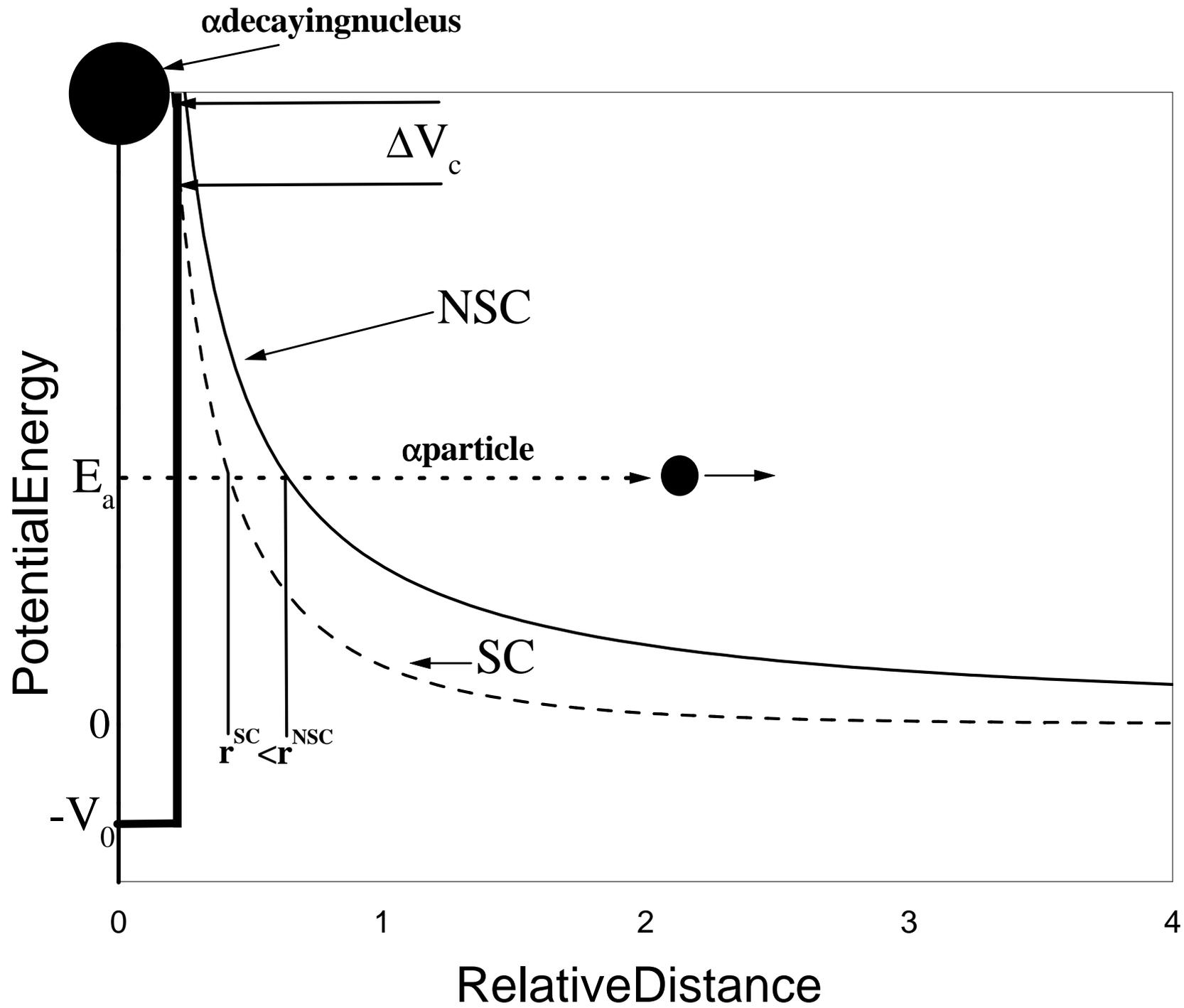
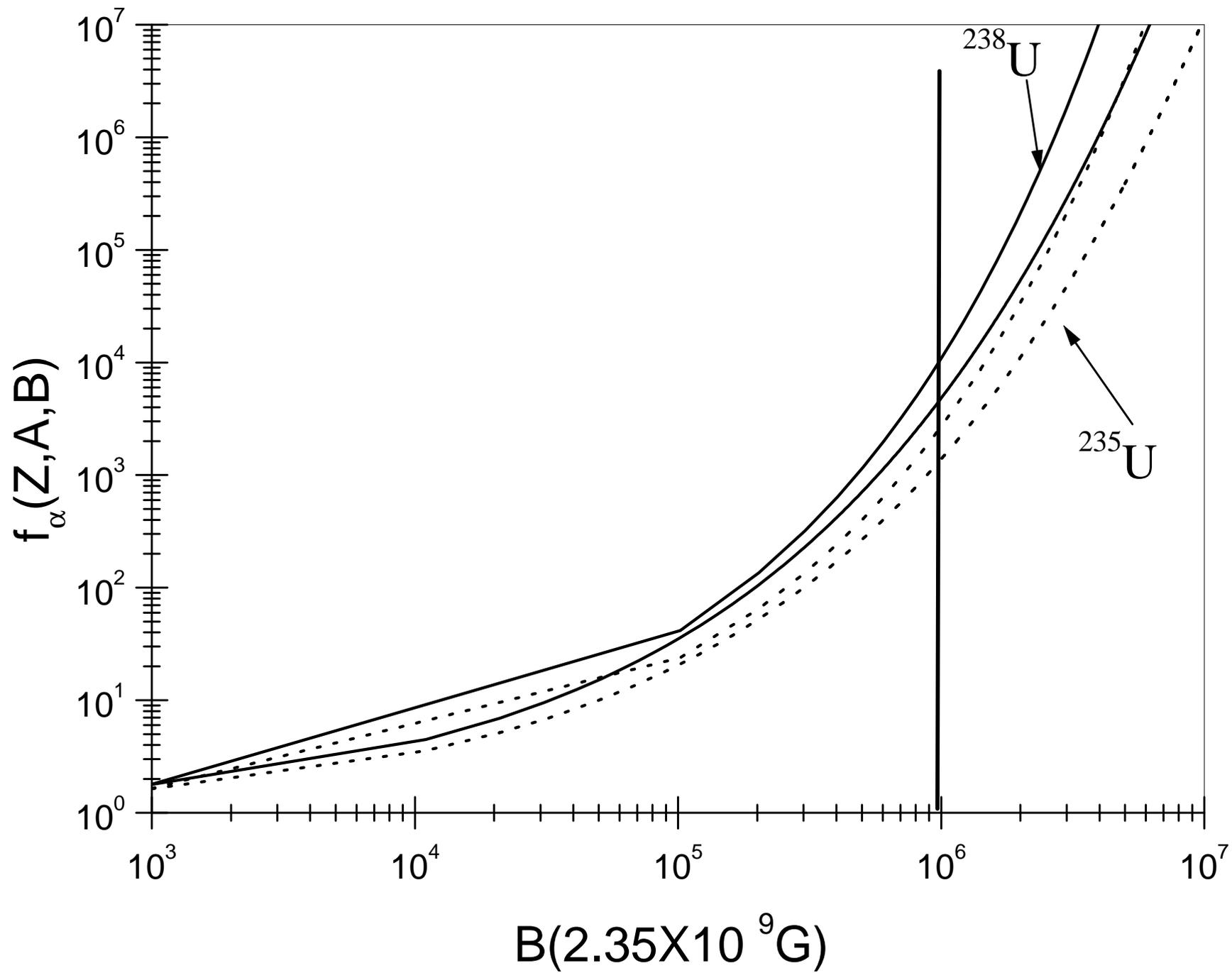
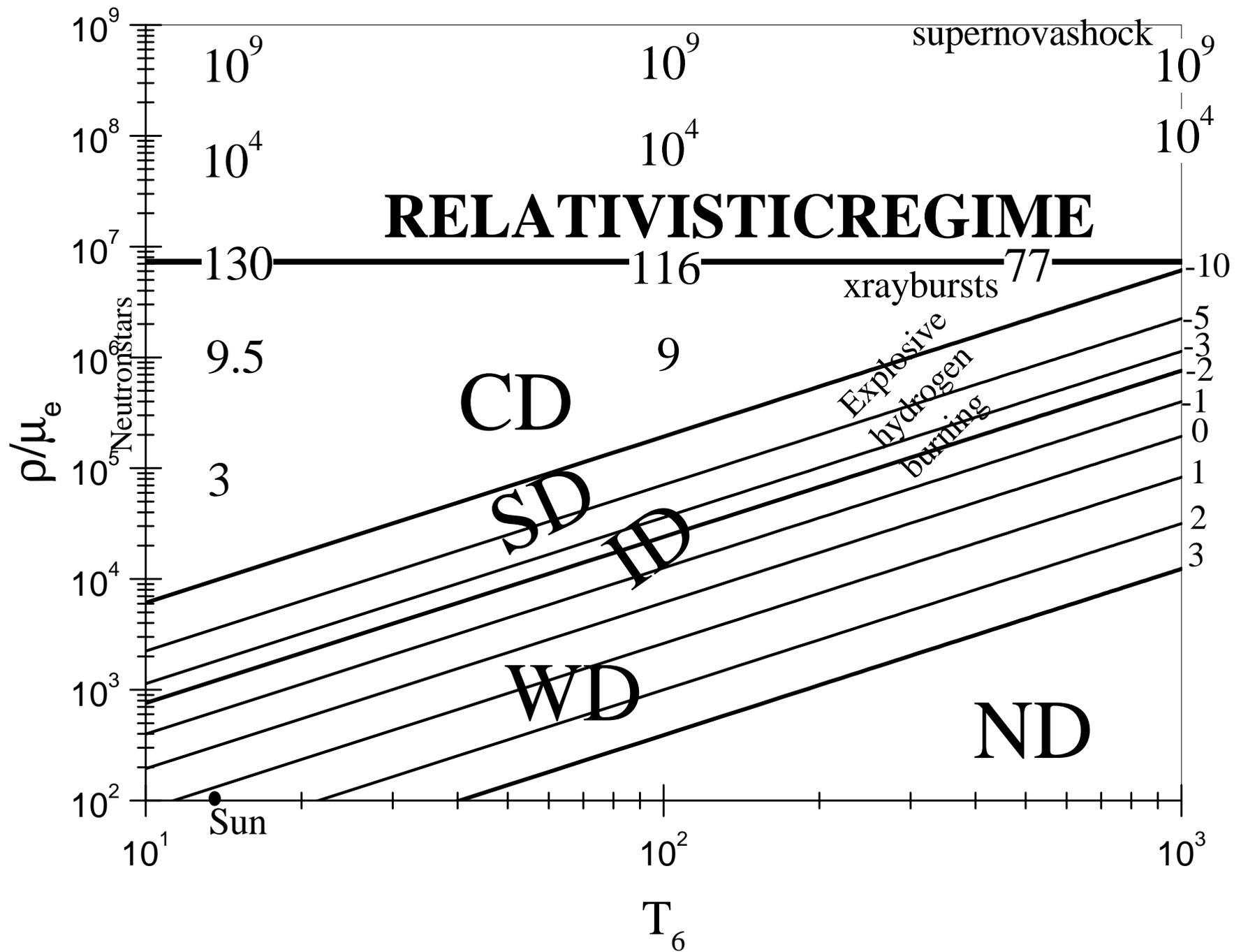
\end{document}